\documentclass[aps,twocolumn,floatfix,showpacs,amsmath,amsfonts]{revtex4}

\usepackage{graphicx}
\usepackage{bm}
\usepackage{units}
\usepackage{color}
\usepackage{upgreek}

\usepackage{blindtext}

\begin{document}

\title{Crossover between the Gaussian orthogonal ensemble, the Gaussian unitary ensemble, and Poissonian statistics}

\author{Frank Schweiner}
\author{Jeanine Laturner}
\author{J\"org Main}
\author{G\"unter Wunner}
\affiliation{Institut f\"ur Theoretische Physik 1, Universit\"at Stuttgart,
  70550 Stuttgart, Germany}
\date{\today}

\begin{abstract}
Until now only for specific crossovers between 
Poissonian statistics (P),
the statistics of a Gaussian orthogonal ensemble (GOE), or the 
statistics of a Gaussian unitary ensemble (GUE)
analytical formulas for the level spacing distribution function 
have been derived within random matrix theory.
We investigate arbitrary crossovers in the triangle between all three statistics.
To this aim we propose an according formula for
the level spacing distribution function depending on two parameters.
Comparing the behavior of our formula for the special cases
of P$\rightarrow$GUE, P$\rightarrow$GOE, and GOE$\rightarrow$GUE with
the results from random matrix theory, we prove that these crossovers are 
described reasonably.
Recent investigations by F.~Schweiner~\emph{et al.}
[Phys. Rev. E~\textbf{95}, 062205 (2017)] have shown that
the Hamiltonian of magnetoexcitons in cubic semiconductors can exhibit
all three statistics in dependence on the system parameters. 
Evaluating the numerical results for magnetoexcitons
in dependence on the excitation energy
and on a parameter connected with the cubic valence band structure
and comparing the results with the formula proposed 
allows us to distinguish between regular and chaotic behavior as well as
between existent or broken antiunitary symmetries. Increasing one of the two
parameters, transitions between different crossovers, e.g., from the
P$\rightarrow$GOE to the P$\rightarrow$GUE-crossover, are observed and discussed.

\end{abstract}


\maketitle

\section{Introduction}

It is now widely accepted that classical chaotic dynamics
manifests itself in the statistical quantities
of the corresponding quantum system~\cite{GUE5,QSC_29,QSC_30}.
All systems with a Hamiltonian leading to 
global chaos in the classical dynamics
can be assigned to one of three universality classes: the orthogonal, the unitary or
the symplectic universality class~\cite{QSC}.
To which of these universality classes a given system
belongs is determined by the remaining symmetries of the system.
Many physical systems are invariant under time-reversal or 
possess at least one remaining antiunitary symmetry.
These systems show the statistics of a Gaussian orthogonal ensemble (GOE).
Only if all antiunitary symmetries are broken,
the statistics of a Gaussian unitary ensemble (GUE) occurs.
The Gaussian symplectic ensemble will not be treated here 
and is described, e.g., in Ref.~\cite{QSC}.
Until now only few physical systems are known
showing a crossover between GOE and GUE statistics in dependence on the system parameters:
the kicked top~\cite{QSC_K7_36}, 
the Anderson model~\cite{GUE4_21}, 
and magnetoexcitons in cubic semiconductors~\cite{175,QC}.
While the kicked top is a time-dependent system, which has to be treated
within Floquet theory~\cite{GUE3,QSC_K7_36}, and 
the Anderson model is rather a model system
for a $d$-dimensional disordered lattice~\cite{GUE4_21},
we showed in Ref.~\cite{225} that magnetoexcitons, 
i.e., excitons in magnetic fields, are a realistic
physical system perfectly suitable to study crossovers between
the Poissonian (P) level statistics, which describes the classically integrable 
case, GOE statistics, and GUE statistics.

Only for the specific crossovers
of P$\rightarrow$GOE, P$\rightarrow$GUE, and GOE$\rightarrow$GUE
analytical formulas for the level spacing distribution function 
have been derived within random matrix theory~\cite{GUE4}.
We have recently investigated the crossovers P$\rightarrow$GUE
and GOE$\rightarrow$GUE for magnetoexcitons~\cite{225} and obtained
a very good agreement with these functions.
However, what has not been investigated so far
are arbitrary crossovers in the triangle between all three statistics
in dependence on two of the system parameters.
In this paper we will investigate these crossovers 
in dependence on the energy 
and one of the Luttinger parameters, which describes the cubic warping
of the valence bands in a semiconductor.
Within random matrix theory it would be, in principle, possible
to derive an analytical formula which describes these arbitrary crossovers
and with which our results for magnetoexcitons could be compared.
However, this derivation is very challenging and beyond the scope of the present work.
On the other hand, crossovers between different symmetry classes are not universal~\cite{GUE10}.
Hence, we propose a function with two parameters for arbitrary crossovers
and show that it describes these crossovers reasonably well
by comparing it for the special cases
of P$\rightarrow$GOE, P$\rightarrow$GUE, and GOE$\rightarrow$GUE
with the analytical formulas known.
We choose the two parameters such that one describes the crossover
from regular to irregular behavior and that the other one describes the
breaking of antiunitary symmetries. Hence, by evaluating the numerical
results with the function proposed, we can 
distinguish between regular and chaotic behavior as well as
between existent or broken antiunitary symmetries.
Varying one of the two
control parameters allows us to 
observe and discuss transitions between different crossovers, e.g., from the
P$\rightarrow$GOE to the P$\rightarrow$GUE-crossover.

\begin{figure*}
\begin{centering}
\includegraphics[width=2.0\columnwidth]{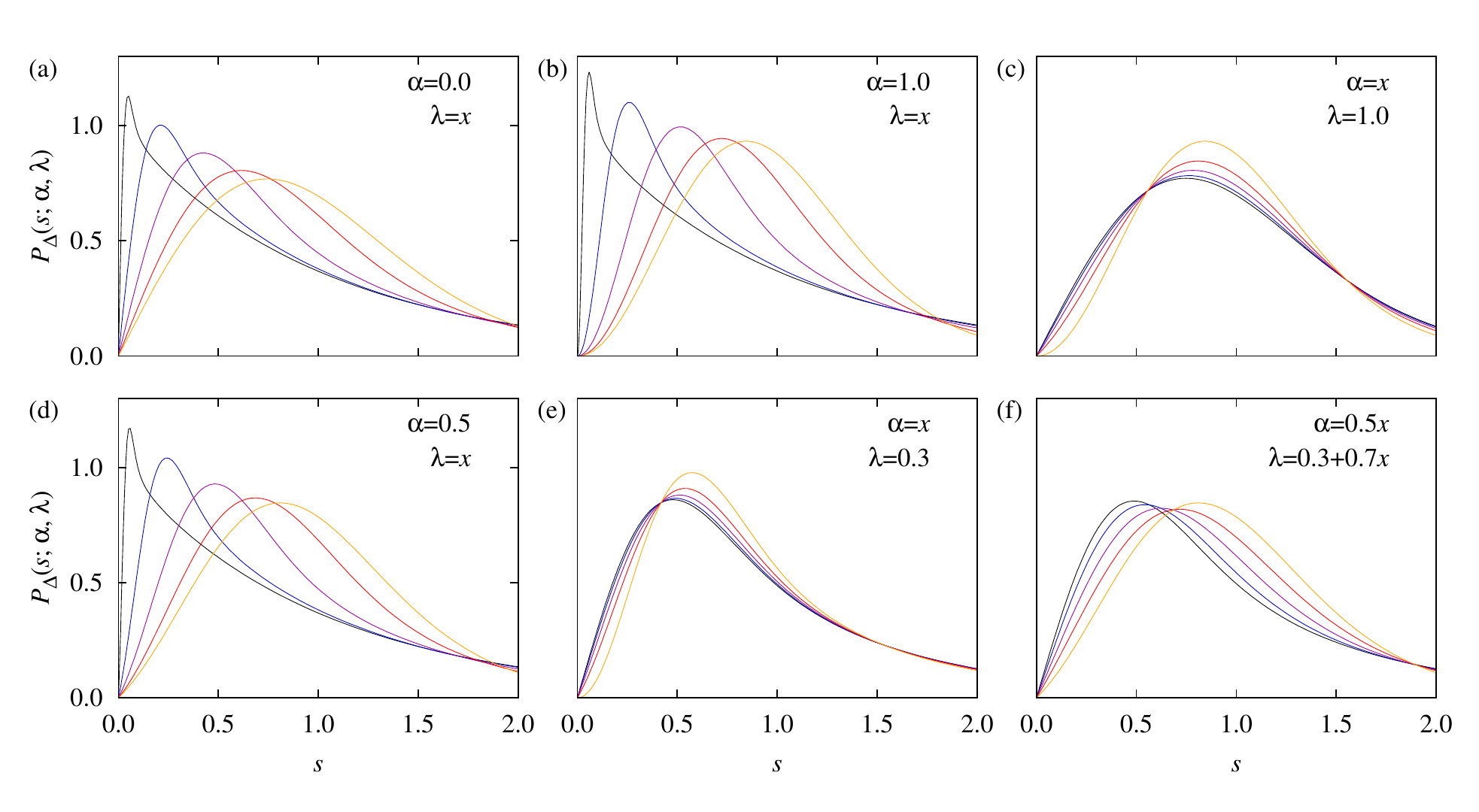}
\par\end{centering}

\protect\caption{The crossover function 
$P_{\triangle}(s;\,\alpha,\,\lambda)$ of Eq.~(\ref{eq:Ptri}) for different
combinations of the parameters $\alpha$ and $\lambda$.
The values of these parameters are given according to the
linear equations in each panel with 
$x=0.02,\,0.10,\,0.25,\,0.5,\,1.0$ (from dark to bright
or left to right).~\label{fig:Ptri}}

\end{figure*}

The paper is organized as follows:
In Sec.~\ref{sec:Theory} we propose the function
for arbitrary crossovers in the triangle P-GOE-GUE and
compare it with the results from random matrix theory for specific crossovers.
After a short discussion of the model system of magneoexcitons 
in cubic semiconductors in Sec.~\ref{sec:magnetoexcitons},
we present a comprehensive discussion
of the numerical results for all possible crossovers in the 
triangle in Sec.~\ref{sec:Results}.
Finally, we give a short summary and outlook in Sec.~\ref{sec:summary}.

\section{Crossover functions\label{sec:Theory}}

In this section we propose a formula
for arbitrary crossovers in the triangle of Poissonian,
GOE and GUE statistics.
For crossovers between each two of the statistics
analytical formulas have been derived within random
matrix theory in Ref.~\cite{GUE4}.
They investigated the statistical properties of a
$2\times 2$ random matrix of the form
\begin{equation}
H=H_{\beta}+\lambda H_{\beta'}
\end{equation}
with a coupling parameter $\lambda$.
$H_{\beta'}$ describes the perturbation breaking
the symmetry of the original system $H_{\beta}$. 
The Poisson process is defined by
\begin{equation}
H_{0}=\left(\begin{array}{cc}
0 & 0\\
0 & p
\end{array}\right)
\end{equation}
with a Poisson-distributed non-negative random number $p$.
The GOE process and the GUE process are described
by a real symmetric matrix
\begin{equation}
H_{1}=\left(\begin{array}{cc}
a & c\\
c & b
\end{array}\right)
\end{equation}
and a complex Hermitian matrix
\begin{equation}
H_{2}=\left(\begin{array}{cc}
a & c_0+ic_1\\
c_0-ic_1 & b
\end{array}\right),
\end{equation}
respectively.
A detailed evaluation of the level spacing distribution
yields the probability densities to find two neighboring
eigenvalues at a distance $s$~\cite{GUE4}:
$P_{\mathrm{P}\rightarrow\mathrm{GOE}}(s;\,\lambda)$,
$P_{\mathrm{P}\rightarrow\mathrm{GUE}}(s;\,\lambda)$, and
$P_{\mathrm{GOE}\rightarrow\mathrm{GUE}}(s;\,\lambda)$.
These formulas are presented in detail in Refs.~\cite{GUE4,225}.
It is important to note that the parameter
$\lambda$ can have all values between $0$ and $\infty$.
However, already for $\lambda\approx 1$ the crossover
to the statistics of lower symmetry is almost completed~\cite{225}.

For the most general case of arbitrary crossovers
between the three processes, one would have to choose
the ansatz
\begin{equation}
H=H_{0}+\lambda_1 H_{1}+\lambda_2 H_{2}
\end{equation}
to derive the nearest-neighbor spacing distribution
$P_{\mathrm{P}-\mathrm{GOE}-\mathrm{GUE}}(s;\,\lambda_1;\,\lambda_2)$.
However, as already the exact analytical calculations
of Ref.~\cite{GUE4} are very complicated, we here
present a different approach.

We already stated in the introduction that the crossover 
between different symmetry classes is not universal.
Besides the crossover formulas derived within 
random matrix theory there are
also other interpolating distributions, e.g., for the crossover 
$\mathrm{P}\rightarrow\mathrm{GOE}$, which have been proposed 
in the literature~\cite{GUE3_10a,GUE3_10b,GUE3_10c,GUE3_10d,GUE3_10e}.
Hence, we also propose a new formula for the arbitrary
crossovers based on the formulas of random matrix theory.
We define the function
\begin{eqnarray}
P_{\triangle}(s;\,\alpha,\,\lambda) & \equiv & P_{\mathrm{P}-\mathrm{GOE}-\mathrm{GUE}}(s;\,\alpha;\,\lambda)\nonumber\\
& = & (1-\alpha) P_{\mathrm{P}\rightarrow\mathrm{GOE}}(s;\,\lambda)\nonumber\\
&  & +\alpha P_{\mathrm{P}\rightarrow\mathrm{GUE}}(s;\,\lambda),\label{eq:Ptri}
\end{eqnarray}
which is normalized
\begin{equation}
\int_{0}^{\infty}\,\mathrm{d}s\,P_{\triangle}(s;\,\alpha,\,\lambda)=(1-\alpha)+\alpha=1
\end{equation}
and fulfils the condition
\begin{equation}
\int_{0}^{\infty}\,\mathrm{d}s\,s P_{\triangle}(s;\,\alpha,\,\lambda)=(1-\alpha)+\alpha=1
\end{equation}
for the mean spacing.
In Fig.~\ref{fig:Ptri} we show the function
$P_{\triangle}(s;\,\alpha,\,\lambda)$ for different
values of $\alpha$ and $\lambda$.

\begin{figure}[t]
\begin{centering}
\includegraphics[width=1.0\columnwidth]{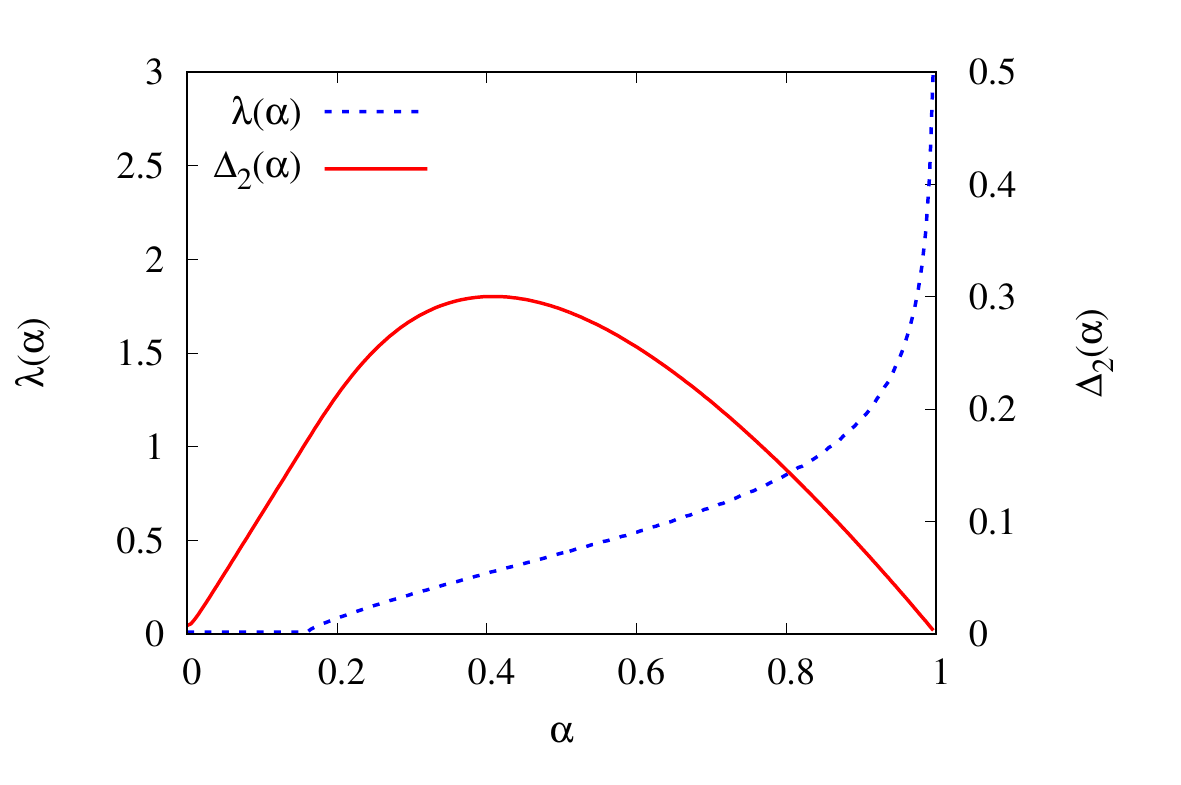}
\par\end{centering}

\protect\caption{The optimum values of the
parameter $\lambda$ when fitting 
$P_{\mathrm{GOE}\rightarrow\mathrm{GUE}}(s;\,\lambda)$
to $P_{\triangle}(s;\,\alpha,\,10)$ for given values of $\alpha$.
With these values the distance $\Delta_2$
has been calculated according to Eq.~(\ref{eq:Delta}). 
For further information see text.~\label{fig:Delta}}

\end{figure}

It can be easily seen that this
function correctly describes the crossovers
P$\rightarrow$GOE (for $\alpha=0$) and P$\rightarrow$GUE (for $\alpha=1$).
When setting $\lambda\gg 1$ and increasing $\alpha$
from $0$ to $1$ this function should also describe the remaining crossover
GOE$\rightarrow$GUE.
Therefore, we fit the function
$P_{\mathrm{GOE}\rightarrow\mathrm{GUE}}(s;\,\lambda)$ from random
matrix theory to $P_{\triangle}(s;\,\alpha,\,10)$ for given values of $\alpha$
using $\lambda$ as a fit parameter (cf.~Refs.~\cite{GUE4,225},
where the maximum value of $\lambda$ is 10).
For the optimum values $\lambda(\alpha)$, we then
calculate the $L_2$ distance
\begin{eqnarray}
\Delta_2(\alpha) & = & \left[\int_{0}^{\infty}\,\mathrm{d}s\,\left[P_{\mathrm{GOE}\rightarrow\mathrm{GUE}}(s;\,\lambda(\alpha))\right.\right.\nonumber\\
& & \qquad\qquad\left.\phantom{\int_{0}^{\infty}}\left.-P_{\triangle}(s;\,\alpha,\,1)\right]^2\right]^{1/2}\label{eq:Delta}
\end{eqnarray}
as a measure of the fit quality~\cite{GUE4}.
The results for $\Delta_2(\alpha)$
and $\lambda(\alpha)$ are shown in Fig.~\ref{fig:Delta}.
It can be seen that the value of $\lambda$ grows monotonically
for increasing values of $\alpha$ and that $\Delta_2(\alpha)$
approaches zero for $\alpha\rightarrow 0$
and $\alpha\rightarrow 1$, which describe the limiting
cases of GOE and GUE statistics, respectively,
Both observations indicate that 
our function $P_{\triangle}(s;\,\alpha,\,10)$
describes the crossover GOE$\rightarrow$GUE reasonably well.
It is understandable that our function
deviates from $P_{\mathrm{GOE}\rightarrow\mathrm{GUE}}(s;\,\lambda)$
for $0<\alpha<1$. 
For $\alpha\approx 0.4$ the deviation is largest 
with $\Delta_2\approx 0.3$. 
Due to these findings and the fact that crossover functions
are not universal, we are certain that the function
$P_{\triangle}(s;\,\alpha,\,\lambda)$ 
provides an adequate description of crossovers in the
triangle of Poisson, GOE, and GUE statistics.

We finally note that the value of the
parameter $\alpha$ in Eq.~(\ref{eq:Ptri}) 
is ambiguous for $\lambda=0$ since it is
\begin{eqnarray}
P_{\triangle}(s;\,\alpha,\,0) & = & (1-\alpha) P_{\mathrm{P}}(s)+\alpha P_{\mathrm{P}}(s)=P_{\mathrm{P}}(s).\label{eq:Ptri}
\end{eqnarray}
Hence, when having fitted the function 
$P_{\triangle}(s;\,\alpha,\,\lambda)$ to numerical results,
we always present the product $\alpha\lambda$ instead of $\alpha$.

\begin{figure}[t]
\begin{centering}
\includegraphics[width=0.68\columnwidth]{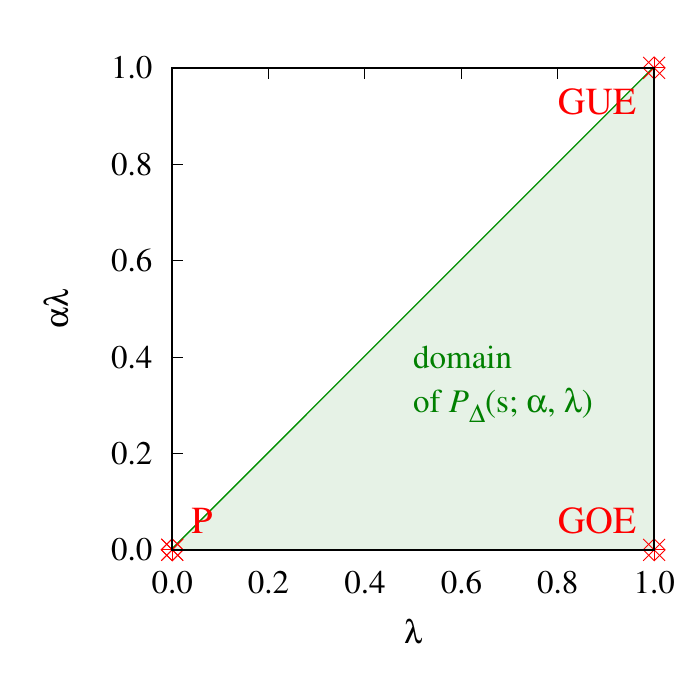}
\par\end{centering}

\protect\caption{The triangle of the different statistics with 
Poissonian $(\lambda=0)$, GOE $(\alpha=0,\,\lambda=1)$, 
and GUE statistics $(\alpha=1,\,\lambda=1)$ located at the corners. 
The green area shows the domain
of the function $P_{\triangle}(s;\,\alpha,\,\lambda)$ .~\label{fig:Pdom}}

\end{figure}

In Fig.~\ref{fig:Pdom} we show the triangle of Poissonian,
GOE, and GUE statistics, which will be important when discussing the
numerical results.
Since we plot $\alpha\lambda$ against $\lambda$, the lower
left corner corresponds to Poissonian statistics
while the lower right corner and the upper right corner
correspond to GOE statistics and GUE statistics, respectively.
The green solid line shows the value of $\alpha=1$.

\section{Magnetoexcitons\label{sec:magnetoexcitons}}

Excitons in semiconductors are fundamental
quasi-particles, which are often regarded as the hydrogen analog
of the solid state. They consist of a negatively charged
electron in the conduction band and a positivley charged
hole in the valence band interacting via a Coulomb interaction
which is screened by the dielectric constant.
Especially for cuprous oxide $\left(\mathrm{Cu_{2}O}\right)$ 
an almost perfect hydrogen-like absorption series
has been observed for the yellow exciton
up to a principal quantum number of $n=25$~\cite{GRE}. 
This remarkable high-resolution absorption experiment
has opened the field of research of giant Rydberg excitons,
and stimulated a large number of experimental and theoretical 
investigations~\cite{GRE,QC,QC2,75,76,50,28,80,100,125,175,78,79,
150,74,77,200,225,275,250,300,70,94,95,96,97}.

When treating excitons in magnetic fields, i.e., magnetoexcitons,
it is indispensable to account for the complete cubic
valence band structure of a semiconductor in a quantitative theory~\cite{125}.
Very recently, we have shown that this cubic valence band structure
breaks all antiunitary symmetries~\cite{175}
and that, depending on the system parameters, Poissonian, GOE and GUE statistics
can be obeserved~\cite{225}.

The Hamiltonian of magnetoexcitons has been discussed thoroughly
in Refs.~\cite{125,225,275}. In this paper we use the simplified model
of magnetoexcitons of Ref.~\cite{225}, in which the spins 
of the electron and the hole are neglected. 
Without the magnetic field the Hamiltonian of the relative motion
between electron and hole reads in terms of irreducible tensors
\begin{eqnarray}
H_0 & = & -\frac{e^{2}}{4\pi\varepsilon_{0}\varepsilon}\frac{1}{r}+ \frac{\gamma'_{1}}{2\hbar^{2}m_{0}}\left[\frac{\delta'}{3}\left(\sum_{k=\pm4}\left[P^{(2)}\times I^{(2)}\right]_{k}^{(4)}\right.\right.\nonumber\\
\nonumber\\
& + & \left.\left.\frac{\sqrt{70}}{5}\left[P^{(2)}\times I^{(2)}\right]_{0}^{(4)}\right)+\hbar^{2}p^{2}-\frac{\mu'}{3}P^{(2)}\cdot I^{(2)}\right]\nonumber\\
\label{P2eq:H0}
\end{eqnarray}
with the dielectric constant $\varepsilon$ and the parameters
$\gamma_1'$, $\mu'$ and $\delta'$, which are connected 
to the Luttinger parameters of the semiconductor
and describe the curvature of the uppermost valence bands~\cite{25,7,100}.
The tensor operators correspond to the Cartesian operators
of the relative momentum $\boldsymbol{p}$ and the quasi-spin $I=1$,
which is connected with the three uppermost valence bands.
The parameter $\delta'$ is of particular importance since
it describes the cubic warping of the valence bands and thus the
breaking of the spherical symmetry of the remaining terms
in the Hamiltonian.
The magnetic field $\boldsymbol{B}$ can finally be introduced in the Hamiltonian $H_0$
via the minimal substitution~\cite{125}.

\begin{figure}[t]
\begin{centering}
\includegraphics[width=1.0\columnwidth]{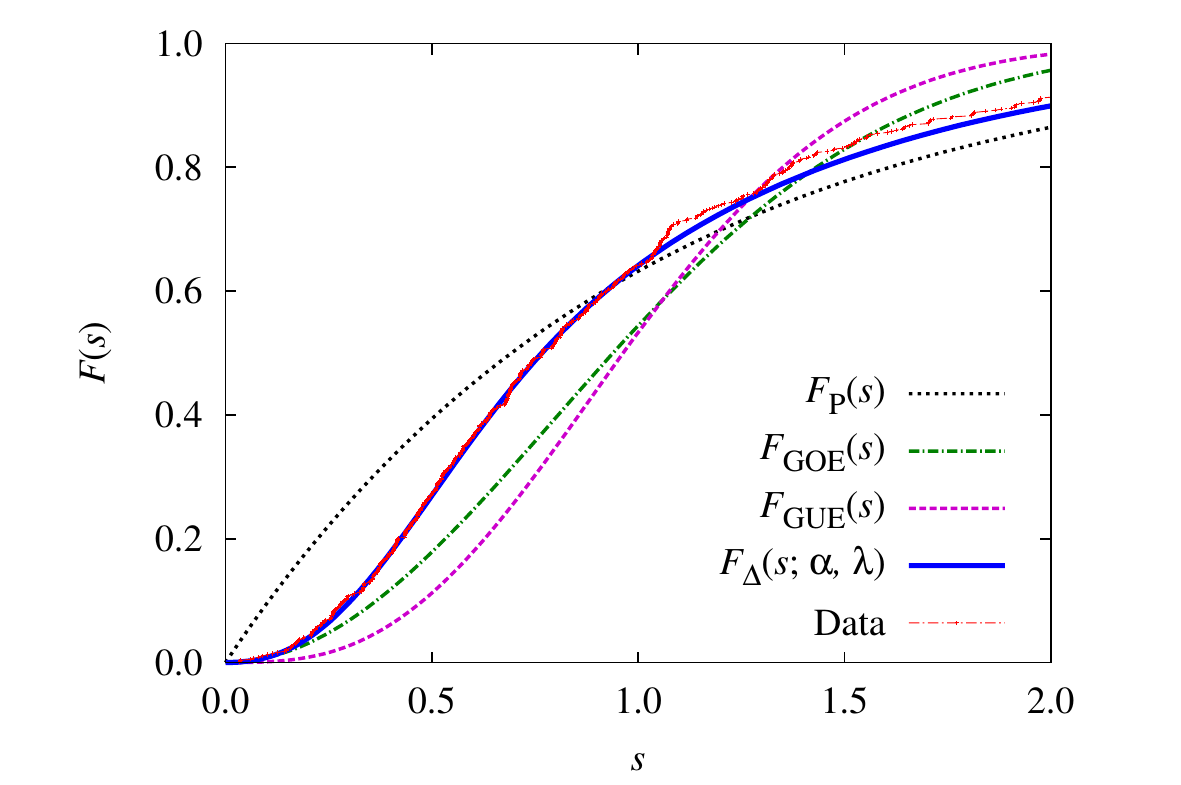}
\par\end{centering}

\protect\caption{Cumulative distribution function
for $\delta'=-0.04$ and $\hat{E}=-0.6$. The numerical data
(red linespoints) is fitted by the cumulative distribution function
$F_{\triangle}(s;\,\alpha,\,\lambda)$ corresponding to the
level spacing distribution of Eq.~(\ref{eq:Ptri}).
The optimum fit parameters are here $\alpha=0.65$ and
$\lambda=0.261$. Hence, the statistics is in the middle
between Poissonian, GOE, and GUE statistics.~\label{fig:eval}}

\end{figure}

We have shown in Refs.~\cite{175,225,275} that if the magnetic field
is not oriented in one of the symmetry planes of the lattice,
all antiunitary symmetries are broken unless $\delta'=0$ holds.
For the subsequent calculations we choose the orientation
of $\boldsymbol{B}$ given by the angles
$\varphi=\pi/8$ and $\vartheta=\pi/6$ in spherical coordinates,
which is far away from the symmetry planes (cf.~Ref.~\cite{225}).

We also use the method of a constant
scaled energy known from atomic physics~\cite{GUE5_23}.
Within this method the coordinate $r$, momentum $p$, and the energy $E$
are scaled by factor a $\gamma=B/B_0$ with
$B_0=2.3505\times 10^5\,\mathrm{T}/(\gamma_1'^2\varepsilon^2)$
as described in detail in Ref.~\cite{225}.
The Schr\"odinger equation can then be written
as a generalized eigenvalue problem
\begin{equation}
\boldsymbol{D}\boldsymbol{c}=\gamma^{1/3}\boldsymbol{M}\boldsymbol{c}
\end{equation}
using the complete basis of Ref.~\cite{225}.
The matrices $\boldsymbol{D}$ and $\boldsymbol{M}$
and, hence, also the solutions of the Schr\"odinger equation 
depend on the two parameters $\hat{E}$ and $\delta'$.
It is well known from atomic physics that for small values
of $\hat{E}$ the behavior of the system is regular while
it becomes chaotic for larger values of $\hat{E}$.
Consequently, $\hat{E}$ and $\delta'$ are the important parameters
when describing arbitrary crossovers in the triangle of
Poissonian, GOE, and GUE statistics. We investigate the 
level spacing statistics of the eigenvalues of the Hamiltonian $H(\delta',\,\hat{E})$ 
depending on these two parameters in the next section~\ref{sec:Results}.

\section{Results and discussion\label{sec:Results}}

Having solved the Schr\"odinger equation
corresponding to the Hamiltonian $H(\delta',\,\hat{E})$
of magnetoexcitons, we unfold the spectra 
according to the descriptions in Ref.~\cite{225} to obtain 
a constant mean spacing~\cite{GUE1,QSC,QC_1,QC_16}.
In doing so, we have to leave out a certain number 
of low-lying sparse levels to remove individual but
nontypical fluctuations~\cite{GUE1}.
Since the number of level spacings analyzed
is comparatively small and comprises
about $250$ to $500$ exciton states, we use
the cumulative distribution function~\cite{GUE2}
\begin{equation}
F(s)=\int_{0}^{s}P(x)\,\mathrm{d}x,
\end{equation}
which is often more meaningful than
histograms of the level
spacing probability distribution function $P(s)$.

\begin{figure}[t]
\begin{centering}
\includegraphics[width=1.0\columnwidth]{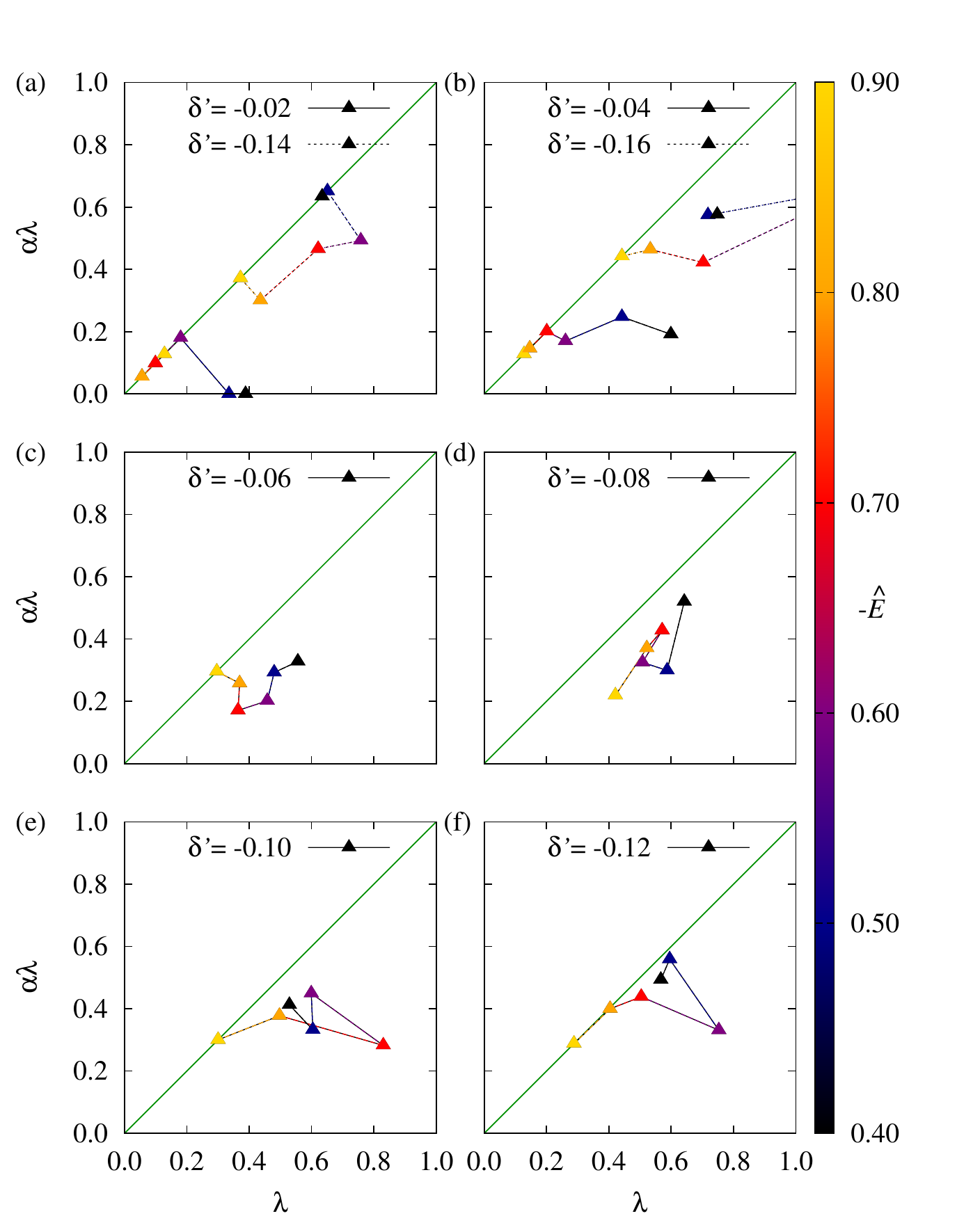}
\par\end{centering}

\protect\caption{Resulting values for the parameters
$\alpha\lambda$ and $\lambda$ when fitting the function
$F_{\triangle}(s;\,\alpha,\,\lambda)$ corresponding to the
level spacing distribution of Eq.~(\ref{eq:Ptri}) to the
cumulative distribution function of the magnetoexciton.
Here we show the behavior of the two fit parameters
when keeping the value $\delta'$ fixed (see label in the panels)
and increasing the scaled energy $\hat{E}$ (color scale).
~\label{fig:Eincr}}

\end{figure}

\begin{figure}[t]
\begin{centering}
\includegraphics[width=1.0\columnwidth]{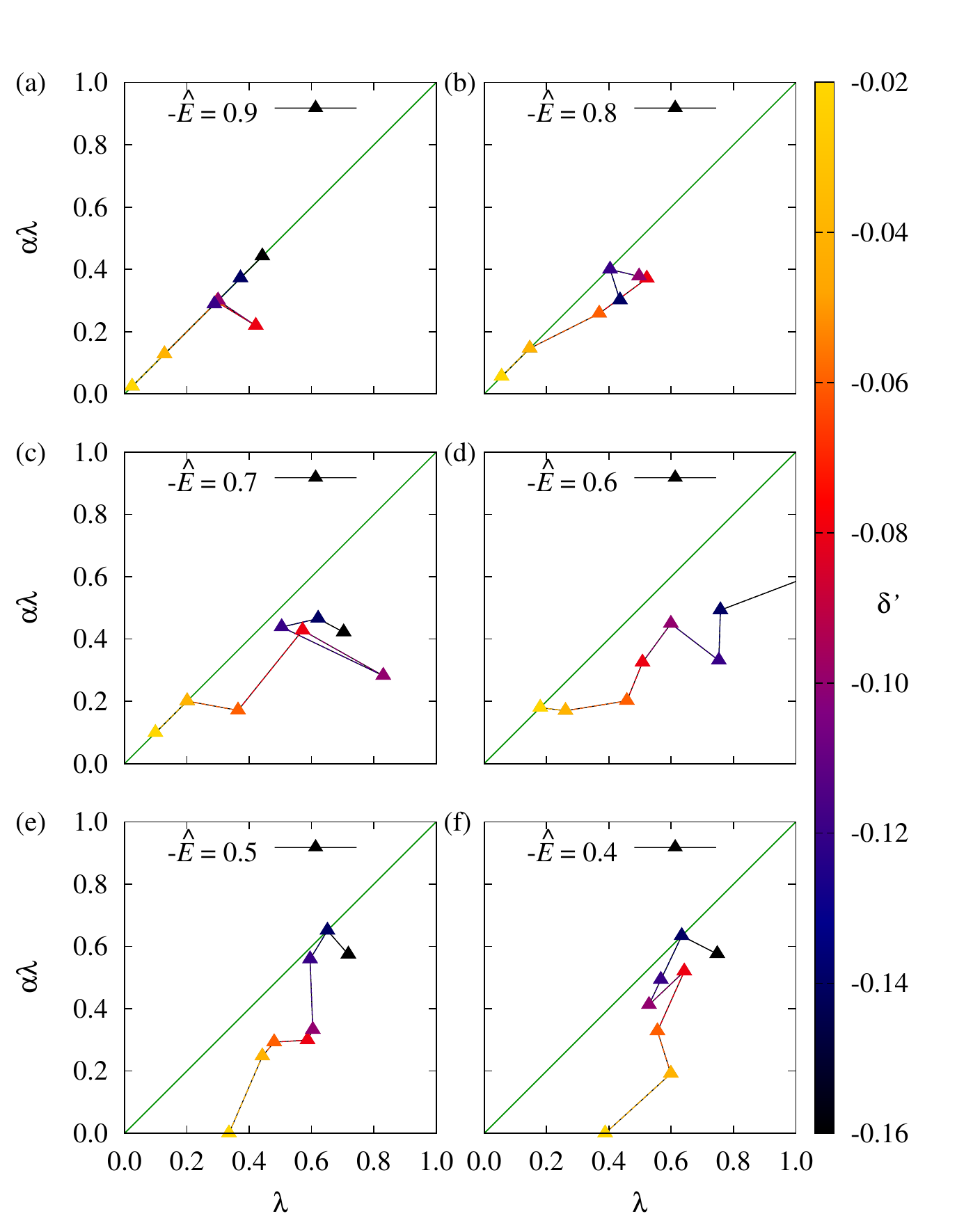}
\par\end{centering}

\protect\caption{Same results as in Fig.~\ref{fig:Eincr}
but shown for fixed values of the scaled energy $\hat{E}$ 
(see label in the panels)
and decreasing values of $\delta'$ (color scale).
~\label{fig:dincr}}

\end{figure}

The numerical results are then fitted by the cumulative distribution 
function $F_{\triangle}(s;\,\alpha,\,\lambda)$ corresponding 
to the level spacing distribution of Eq.~(\ref{eq:Ptri}).
This is shown exemplarily in Fig.~\ref{fig:eval}.
As can be seen, the
agreement between the results and the function $F_{\triangle}(s;\,\alpha,\,\lambda)$
is reasonable. Note that this is generally true for all parameter sets.
We evaluate numerical spectra for $\delta'=-0.02,\,-0.04,\,\ldots,\,-0.16$
and $\hat{E}=-0.4,\,-0.5,\,\ldots,\,-0.9$. 

The results for the fit parameters $\alpha$
and $\lambda$ are shown in Figs.~\ref{fig:Eincr} and~\ref{fig:dincr}.
The two figures show the change in the fit parameters when
keeping one of the two values $\delta'$ and $\hat{E}$ fixed
and varying the other one.

Let us start with Fig.~\ref{fig:Eincr}
and the evaluation for fixed values of the parameter $\delta'$.
In the limit $\delta'\rightarrow 0$ the influence
of the cubic valence band structure vanishes and
the system becomes hydrogen-like.
It is well known that the hydrogen atom shows Poissonian
statistics for small values of $\hat{E}$ and that a crossover
to GOE statistics occurs when increasing the scaled energy~\cite{GUE1}.
Hence, we expect for very small values of $|\delta'|$ an almost horizontal
line in the figures at small values of $\alpha\lambda$. This can be seen
in Fig.~\ref{fig:Eincr} for $\delta'=-0.02$ and even better for $\delta'=-0.04$.

Here we already want to state that due to the comparatively
small number of exciton states, which can be used
in the numerical evaluation, the numerical data shows
some fluctuations as can be seen, e.g., for larger values of $s$ in 
Fig.~\ref{fig:eval}. Furthermore, when varying the parameters
$\alpha$ and $\lambda$ only slightly, the shape of the function 
$P_{\triangle}(s;\,\alpha,\,\lambda)$ or 
$F_{\triangle}(s;\,\alpha,\,\lambda)$ hardly changes 
(cf.~also Fig.~\ref{fig:Ptri}).
Consequently, due to these facts the results shown in 
Figs.~\ref{fig:Eincr} and~\ref{fig:dincr} 
also show some fluctuations. However, one can nevertheless
see the general behavior, when changing the $\delta'$ and $\hat{E}$.

When increasing $|\delta'|$ the cubic valence band structure becomes important
and all antiunitary symmetries are broken. Hence, we see from Fig.~\ref{fig:Eincr}
that the points are shifted towards higher values of $\alpha\lambda$
indicating that the line statistics becomes more and more GUE-like.
We also observe that the line for fixed values of $\delta'$ tends
to change its shape from an almost horizontal line to a more
diagonal line. The crossover for fixed values of 
$\delta'$ and increasing $\hat{E}$ becomes more and more
P$\rightarrow$GUE-like as expected.

Let us now turn to Fig.~\ref{fig:dincr}. 
We already observed in Ref.~\cite{225} that the parameter $\delta'$
does not only break the remaining antiunitary symmetry of the
hydrogen atom in external fields but also increases the chaotic behavior.
When keeping the scaled energy $\hat{E}$ fixed at a very
small value $\hat{E}=-0.9$ and increasing $\delta'$, 
the statistics does not remain
Poisson-like but the value of $\lambda$
already increases. Since $\alpha$ furthermore remains constant
with $\alpha=1$, this indicates the crossover from Poissonian
to GUE statistics.
On the other hand, it is known from the
hydrogen atom in external fields that when increasing $\hat{E}$
the behavior of the system becomes more and more chaotic, as well.
For large values of $\hat{E}$ the system stays completely in the
chaotic regime independent of the value of $\delta'$.
This can be seen in Fig.~\ref{fig:dincr} for $\hat{E}=-0.4$,
where the value of $\lambda$ is always larger than $0.4$.
For $\hat{E}\geq-0.4$ the statistics is GOE-like in the hydrogen-like
case with $\delta'\rightarrow 0$. When increasing the value of $|\delta'|$
it becomes more and more GUE-like as expected from the results of Ref.~\cite{175,225}.
Hence, we observe the crossover GOE$\rightarrow$GUE as an almost vertical
line in the lower right panel of Fig.~\ref{fig:dincr}.
For the intermediate values $-0.9\leq\hat{E}\leq-0.4$ of the scaled
energy we observe the transition from the P$\rightarrow$GUE-crossover 
to the GOE$\rightarrow$GUE-crossover as a change in the lineshape
from a diagonal to a more and more vertical line.

\section{Summary and outlook\label{sec:summary}}

We have proposed a new nearest-neighbor spacing distribution
function, which allows to investigate arbitrary crossovers
in the triangle of Poissonian, GOE, and GUE statistics.
Comparing the behavior of this function for the special
cases of P$\rightarrow$GOE, P$\rightarrow$GUE, and GOE$\rightarrow$GUE
with the analytical formulas from random matrix theory,
we could show that our function allows for a reasonable description
of these crossovers.
As excitons in external magnetic fields show all these statistics
in dependence on the system parameters, they are ideally suited
to investigate arbitrary crossovers between the three statistics.
Evaluating numerical spectra for different values of the parameter 
$\delta'$ and the scaled energy $\hat{E}$ we could observe
transitions from the P$\rightarrow$GOE-crossover 
to the P$\rightarrow$GUE-crossover when increasing $\delta'$
or from the P$\rightarrow$GUE-crossover 
to the GOE$\rightarrow$GUE-crossover when increasing $\hat{E}$.

\acknowledgments
F.S.~is  grateful  for  support  from  the
Landesgraduiertenf\"orderung of the Land Baden-W\"urttemberg.

\appendix



\begin{thebibliography}{47}
\expandafter\ifx\csname natexlab\endcsname\relax\def\natexlab#1{#1}\fi
\expandafter\ifx\csname bibnamefont\endcsname\relax
  \def\bibnamefont#1{#1}\fi
\expandafter\ifx\csname bibfnamefont\endcsname\relax
  \def\bibfnamefont#1{#1}\fi
\expandafter\ifx\csname citenamefont\endcsname\relax
  \def\citenamefont#1{#1}\fi
\expandafter\ifx\csname url\endcsname\relax
  \def\url#1{\texttt{#1}}\fi
\expandafter\ifx\csname urlprefix\endcsname\relax\def\urlprefix{URL }\fi
\providecommand{\bibinfo}[2]{#2}
\providecommand{\eprint}[2][]{\url{#2}}

\bibitem[{\citenamefont{Rao and Taylor}(2002)}]{GUE5}
\bibinfo{author}{\bibfnamefont{J.}~\bibnamefont{Rao}} \bibnamefont{and}
  \bibinfo{author}{\bibfnamefont{K.~T.} \bibnamefont{Taylor}},
  \bibinfo{journal}{J. Phys. B: At. Mol. Opt. Phys.}
  \textbf{\bibinfo{volume}{35}}, \bibinfo{pages}{2627} (\bibinfo{year}{2002}).

\bibitem[{\citenamefont{Mehta}(2004)}]{QSC_29}
\bibinfo{author}{\bibfnamefont{M.~L.} \bibnamefont{Mehta}},
  \emph{\bibinfo{title}{Random Matrices}} (\bibinfo{publisher}{Elsevier},
  \bibinfo{address}{Amsterdam}, \bibinfo{year}{2004}), \bibinfo{edition}{3rd}
  ed.

\bibitem[{\citenamefont{Porter}(1965)}]{QSC_30}
\bibinfo{editor}{\bibfnamefont{C.~E.} \bibnamefont{Porter}}, ed.,
  \emph{\bibinfo{title}{Statistical Theory of Spectra}}
  (\bibinfo{publisher}{Academic Press}, \bibinfo{address}{New York},
  \bibinfo{year}{1965}).

\bibitem[{\citenamefont{Haake}(2010)}]{QSC}
\bibinfo{author}{\bibfnamefont{F.}~\bibnamefont{Haake}},
  \emph{\bibinfo{title}{Quantum Signatures of Chaos}}, Springer Series in
  Synergetics (\bibinfo{publisher}{Springer}, \bibinfo{address}{Heidelberg},
  \bibinfo{year}{2010}), \bibinfo{edition}{3rd} ed.

\bibitem[{\citenamefont{Haake et~al.}(1987)\citenamefont{Haake, Ku\'{s}, and
  Scharf}}]{QSC_K7_36}
\bibinfo{author}{\bibfnamefont{H.}~\bibnamefont{Haake}},
  \bibinfo{author}{\bibfnamefont{M.}~\bibnamefont{Ku\'{s}}}, \bibnamefont{and}
  \bibinfo{author}{\bibfnamefont{R.}~\bibnamefont{Scharf}},
  \bibinfo{journal}{Z. Phys. B} \textbf{\bibinfo{volume}{65}},
  \bibinfo{pages}{381} (\bibinfo{year}{1987}).

\bibitem[{\citenamefont{Shukla}(2005)}]{GUE4_21}
\bibinfo{author}{\bibfnamefont{P.}~\bibnamefont{Shukla}}, \bibinfo{journal}{J.
  Phys. Condens. Matter} \textbf{\bibinfo{volume}{17}}, \bibinfo{pages}{1653}
  (\bibinfo{year}{2005}).

\bibitem[{\citenamefont{Schweiner
  et~al.}(2017{\natexlab{a}})\citenamefont{Schweiner, Main, and Wunner}}]{175}
\bibinfo{author}{\bibfnamefont{F.}~\bibnamefont{Schweiner}},
  \bibinfo{author}{\bibfnamefont{J.}~\bibnamefont{Main}}, \bibnamefont{and}
  \bibinfo{author}{\bibfnamefont{G.}~\bibnamefont{Wunner}},
  \bibinfo{journal}{Phys. Rev. Lett.} \textbf{\bibinfo{volume}{118}},
  \bibinfo{pages}{046401} (\bibinfo{year}{2017}{\natexlab{a}}).

\bibitem[{\citenamefont{A{\ss}mann et~al.}(2016)\citenamefont{A{\ss}mann,
  Thewes, Fr\"{o}hlich, and Bayer}}]{QC}
\bibinfo{author}{\bibfnamefont{M.}~\bibnamefont{A{\ss}mann}},
  \bibinfo{author}{\bibfnamefont{J.}~\bibnamefont{Thewes}},
  \bibinfo{author}{\bibfnamefont{D.}~\bibnamefont{Fr\"{o}hlich}},
  \bibnamefont{and} \bibinfo{author}{\bibfnamefont{M.}~\bibnamefont{Bayer}},
  \bibinfo{journal}{Nature Mater.} \textbf{\bibinfo{volume}{15}},
  \bibinfo{pages}{741} (\bibinfo{year}{2016}).

\bibitem[{\citenamefont{Lenz and Haake}(1991)}]{GUE3}
\bibinfo{author}{\bibfnamefont{G.}~\bibnamefont{Lenz}} \bibnamefont{and}
  \bibinfo{author}{\bibfnamefont{F.}~\bibnamefont{Haake}},
  \bibinfo{journal}{Phys. Rev. Lett.} \textbf{\bibinfo{volume}{67}},
  \bibinfo{pages}{1} (\bibinfo{year}{1991}).

\bibitem[{\citenamefont{Schweiner
  et~al.}(2017{\natexlab{b}})\citenamefont{Schweiner, Main, and Wunner}}]{225}
\bibinfo{author}{\bibfnamefont{F.}~\bibnamefont{Schweiner}},
  \bibinfo{author}{\bibfnamefont{J.}~\bibnamefont{Main}}, \bibnamefont{and}
  \bibinfo{author}{\bibfnamefont{G.}~\bibnamefont{Wunner}},
  \bibinfo{journal}{Phys. Rev. E} \textbf{\bibinfo{volume}{95}},
  \bibinfo{pages}{062205} (\bibinfo{year}{2017}{\natexlab{b}}).

\bibitem[{\citenamefont{Schierenberg et~al.}(2012)\citenamefont{Schierenberg,
  Bruckmann, and Wettig}}]{GUE4}
\bibinfo{author}{\bibfnamefont{S.}~\bibnamefont{Schierenberg}},
  \bibinfo{author}{\bibfnamefont{F.}~\bibnamefont{Bruckmann}},
  \bibnamefont{and} \bibinfo{author}{\bibfnamefont{T.}~\bibnamefont{Wettig}},
  \bibinfo{journal}{Phys. Rev. E} \textbf{\bibinfo{volume}{85}},
  \bibinfo{pages}{061130} (\bibinfo{year}{2012}).

\bibitem[{\citenamefont{Kunstman et~al.}(1997)\citenamefont{Kunstman,
  \.{Z}yczkowski, and Zakrzewski}}]{GUE10}
\bibinfo{author}{\bibfnamefont{P.}~\bibnamefont{Kunstman}},
  \bibinfo{author}{\bibfnamefont{K.}~\bibnamefont{\.{Z}yczkowski}},
  \bibnamefont{and}
  \bibinfo{author}{\bibfnamefont{J.}~\bibnamefont{Zakrzewski}},
  \bibinfo{journal}{Phys. Rev. E} \textbf{\bibinfo{volume}{55}},
  \bibinfo{pages}{2446} (\bibinfo{year}{1997}).

\bibitem[{\citenamefont{Berry and Robnik}(1984)}]{GUE3_10a}
\bibinfo{author}{\bibfnamefont{M.~V.} \bibnamefont{Berry}} \bibnamefont{and}
  \bibinfo{author}{\bibfnamefont{M.}~\bibnamefont{Robnik}},
  \bibinfo{journal}{J. Phys. A} \textbf{\bibinfo{volume}{17}},
  \bibinfo{pages}{2413} (\bibinfo{year}{1984}).

\bibitem[{\citenamefont{Brody}(1973)}]{GUE3_10b}
\bibinfo{author}{\bibfnamefont{T.~A.} \bibnamefont{Brody}},
  \bibinfo{journal}{Lett. Nuovo Cimento} \textbf{\bibinfo{volume}{7}},
  \bibinfo{pages}{482} (\bibinfo{year}{1973}).

\bibitem[{\citenamefont{Caurier et~al.}(1990)\citenamefont{Caurier,
  Grammaticos, and Ramani}}]{GUE3_10c}
\bibinfo{author}{\bibfnamefont{E.}~\bibnamefont{Caurier}},
  \bibinfo{author}{\bibfnamefont{B.}~\bibnamefont{Grammaticos}},
  \bibnamefont{and} \bibinfo{author}{\bibfnamefont{A.}~\bibnamefont{Ramani}},
  \bibinfo{journal}{J. Phys. A} \textbf{\bibinfo{volume}{23}},
  \bibinfo{pages}{4903} (\bibinfo{year}{1990}).

\bibitem[{\citenamefont{Hasegawa et~al.}(1988)\citenamefont{Hasegawa, Mikeska,
  and Frahm}}]{GUE3_10d}
\bibinfo{author}{\bibfnamefont{H.}~\bibnamefont{Hasegawa}},
  \bibinfo{author}{\bibfnamefont{H.~J.} \bibnamefont{Mikeska}},
  \bibnamefont{and} \bibinfo{author}{\bibfnamefont{H.}~\bibnamefont{Frahm}},
  \bibinfo{journal}{Phys. Rev. A} \textbf{\bibinfo{volume}{38}},
  \bibinfo{pages}{395} (\bibinfo{year}{1988}).

\bibitem[{\citenamefont{Izrailev}(1990)}]{GUE3_10e}
\bibinfo{author}{\bibfnamefont{F.}~\bibnamefont{Izrailev}},
  \bibinfo{journal}{Phys. Rep.} \textbf{\bibinfo{volume}{5-6}},
  \bibinfo{pages}{299} (\bibinfo{year}{1990}).

\bibitem[{\citenamefont{Kazimierczuk et~al.}(2014)\citenamefont{Kazimierczuk,
  Fr\"{o}hlich, Scheel, Stolz, and Bayer}}]{GRE}
\bibinfo{author}{\bibfnamefont{T.}~\bibnamefont{Kazimierczuk}},
  \bibinfo{author}{\bibfnamefont{D.}~\bibnamefont{Fr\"{o}hlich}},
  \bibinfo{author}{\bibfnamefont{S.}~\bibnamefont{Scheel}},
  \bibinfo{author}{\bibfnamefont{H.}~\bibnamefont{Stolz}}, \bibnamefont{and}
  \bibinfo{author}{\bibfnamefont{M.}~\bibnamefont{Bayer}},
  \bibinfo{journal}{Nature} \textbf{\bibinfo{volume}{514}},
  \bibinfo{pages}{343} (\bibinfo{year}{2014}).

\bibitem[{\citenamefont{Freitag et~al.}(2017)\citenamefont{Freitag,
  Heck\"{o}tter, Bayer, and A{\ss}mann}}]{QC2}
\bibinfo{author}{\bibfnamefont{M.}~\bibnamefont{Freitag}},
  \bibinfo{author}{\bibfnamefont{J.}~\bibnamefont{Heck\"{o}tter}},
  \bibinfo{author}{\bibfnamefont{M.}~\bibnamefont{Bayer}}, \bibnamefont{and}
  \bibinfo{author}{\bibfnamefont{M.}~\bibnamefont{A{\ss}mann}},
  \bibinfo{journal}{Phys. Rev. B} \textbf{\bibinfo{volume}{95}},
  \bibinfo{pages}{155204} (\bibinfo{year}{2017}).

\bibitem[{\citenamefont{Schweiner
  et~al.}(2016{\natexlab{a}})\citenamefont{Schweiner, Main, and Wunner}}]{75}
\bibinfo{author}{\bibfnamefont{F.}~\bibnamefont{Schweiner}},
  \bibinfo{author}{\bibfnamefont{J.}~\bibnamefont{Main}}, \bibnamefont{and}
  \bibinfo{author}{\bibfnamefont{G.}~\bibnamefont{Wunner}},
  \bibinfo{journal}{Phys. Rev. B} \textbf{\bibinfo{volume}{93}},
  \bibinfo{pages}{085203} (\bibinfo{year}{2016}{\natexlab{a}}).

\bibitem[{\citenamefont{Gr\"unwald et~al.}(2016)\citenamefont{Gr\"unwald,
  A{\ss}mann, Heck\"{o}tter, Fr\"{o}hlich, Bayer, Stolz, and Scheel}}]{76}
\bibinfo{author}{\bibfnamefont{P.}~\bibnamefont{Gr\"unwald}},
  \bibinfo{author}{\bibfnamefont{M.}~\bibnamefont{A{\ss}mann}},
  \bibinfo{author}{\bibfnamefont{J.}~\bibnamefont{Heck\"{o}tter}},
  \bibinfo{author}{\bibfnamefont{D.}~\bibnamefont{Fr\"{o}hlich}},
  \bibinfo{author}{\bibfnamefont{M.}~\bibnamefont{Bayer}},
  \bibinfo{author}{\bibfnamefont{H.}~\bibnamefont{Stolz}}, \bibnamefont{and}
  \bibinfo{author}{\bibfnamefont{S.}~\bibnamefont{Scheel}},
  \bibinfo{journal}{Phys. Rev. Lett.} \textbf{\bibinfo{volume}{117}},
  \bibinfo{pages}{133003} (\bibinfo{year}{2016}).

\bibitem[{\citenamefont{Feldmaier et~al.}(2016)\citenamefont{Feldmaier, Main,
  Schweiner, Cartarius, and Wunner}}]{50}
\bibinfo{author}{\bibfnamefont{M.}~\bibnamefont{Feldmaier}},
  \bibinfo{author}{\bibfnamefont{J.}~\bibnamefont{Main}},
  \bibinfo{author}{\bibfnamefont{F.}~\bibnamefont{Schweiner}},
  \bibinfo{author}{\bibfnamefont{H.}~\bibnamefont{Cartarius}},
  \bibnamefont{and} \bibinfo{author}{\bibfnamefont{G.}~\bibnamefont{Wunner}},
  \bibinfo{journal}{J. Phys. B: At. Mol. Opt. Phys.}
  \textbf{\bibinfo{volume}{49}}, \bibinfo{pages}{144002}
  (\bibinfo{year}{2016}).

\bibitem[{\citenamefont{Thewes et~al.}(2015)\citenamefont{Thewes,
  Heck\"{o}tter, Kazimierczuk, A{\ss}mann, Fr\"{o}hlich, Bayer, Semina, and
  Glazov}}]{28}
\bibinfo{author}{\bibfnamefont{J.}~\bibnamefont{Thewes}},
  \bibinfo{author}{\bibfnamefont{J.}~\bibnamefont{Heck\"{o}tter}},
  \bibinfo{author}{\bibfnamefont{T.}~\bibnamefont{Kazimierczuk}},
  \bibinfo{author}{\bibfnamefont{M.}~\bibnamefont{A{\ss}mann}},
  \bibinfo{author}{\bibfnamefont{D.}~\bibnamefont{Fr\"{o}hlich}},
  \bibinfo{author}{\bibfnamefont{M.}~\bibnamefont{Bayer}},
  \bibinfo{author}{\bibfnamefont{M.~A.} \bibnamefont{Semina}},
  \bibnamefont{and} \bibinfo{author}{\bibfnamefont{M.~M.}
  \bibnamefont{Glazov}}, \bibinfo{journal}{Phys. Rev. Lett.}
  \textbf{\bibinfo{volume}{115}}, \bibinfo{pages}{027402}
  (\bibinfo{year}{2015}), \bibinfo{note}{and Supplementary Material}.

\bibitem[{\citenamefont{Sch\"{o}ne et~al.}(2016)\citenamefont{Sch\"{o}ne,
  Kr\"{u}ger, Gr\"{u}nwald, Stolz, Scheel, A{\ss}mann, Heck\"{o}tter, Thewes,
  Fr\"{o}hlich, and Bayer}}]{80}
\bibinfo{author}{\bibfnamefont{F.}~\bibnamefont{Sch\"{o}ne}},
  \bibinfo{author}{\bibfnamefont{S.~O.} \bibnamefont{Kr\"{u}ger}},
  \bibinfo{author}{\bibfnamefont{P.}~\bibnamefont{Gr\"{u}nwald}},
  \bibinfo{author}{\bibfnamefont{H.}~\bibnamefont{Stolz}},
  \bibinfo{author}{\bibfnamefont{S.}~\bibnamefont{Scheel}},
  \bibinfo{author}{\bibfnamefont{M.}~\bibnamefont{A{\ss}mann}},
  \bibinfo{author}{\bibfnamefont{J.}~\bibnamefont{Heck\"{o}tter}},
  \bibinfo{author}{\bibfnamefont{J.}~\bibnamefont{Thewes}},
  \bibinfo{author}{\bibfnamefont{D.}~\bibnamefont{Fr\"{o}hlich}},
  \bibnamefont{and} \bibinfo{author}{\bibfnamefont{M.}~\bibnamefont{Bayer}},
  \bibinfo{journal}{Phys. Rev. B} \textbf{\bibinfo{volume}{93}},
  \bibinfo{pages}{075203} (\bibinfo{year}{2016}).

\bibitem[{\citenamefont{Schweiner
  et~al.}(2016{\natexlab{b}})\citenamefont{Schweiner, Main, Feldmaier, Wunner,
  and Uihlein}}]{100}
\bibinfo{author}{\bibfnamefont{F.}~\bibnamefont{Schweiner}},
  \bibinfo{author}{\bibfnamefont{J.}~\bibnamefont{Main}},
  \bibinfo{author}{\bibfnamefont{M.}~\bibnamefont{Feldmaier}},
  \bibinfo{author}{\bibfnamefont{G.}~\bibnamefont{Wunner}}, \bibnamefont{and}
  \bibinfo{author}{\bibfnamefont{{\relax Ch}.}~\bibnamefont{Uihlein}},
  \bibinfo{journal}{Phys. Rev. B} \textbf{\bibinfo{volume}{93}},
  \bibinfo{pages}{195203} (\bibinfo{year}{2016}{\natexlab{b}}).

\bibitem[{\citenamefont{Schweiner
  et~al.}(2017{\natexlab{c}})\citenamefont{Schweiner, Main, Wunner, Freitag,
  Heck\"{o}tter, Uihlein, A{\ss}mann, Fr\"{o}hlich, and Bayer}}]{125}
\bibinfo{author}{\bibfnamefont{F.}~\bibnamefont{Schweiner}},
  \bibinfo{author}{\bibfnamefont{J.}~\bibnamefont{Main}},
  \bibinfo{author}{\bibfnamefont{G.}~\bibnamefont{Wunner}},
  \bibinfo{author}{\bibfnamefont{M.}~\bibnamefont{Freitag}},
  \bibinfo{author}{\bibfnamefont{J.}~\bibnamefont{Heck\"{o}tter}},
  \bibinfo{author}{\bibfnamefont{{\relax Ch}.}~\bibnamefont{Uihlein}},
  \bibinfo{author}{\bibfnamefont{M.}~\bibnamefont{A{\ss}mann}},
  \bibinfo{author}{\bibfnamefont{D.}~\bibnamefont{Fr\"{o}hlich}},
  \bibnamefont{and} \bibinfo{author}{\bibfnamefont{M.}~\bibnamefont{Bayer}},
  \bibinfo{journal}{Phys. Rev. B} \textbf{\bibinfo{volume}{95}},
  \bibinfo{pages}{035202} (\bibinfo{year}{2017}{\natexlab{c}}).

\bibitem[{\citenamefont{Heck\"otter
  et~al.}(2017{\natexlab{a}})\citenamefont{Heck\"otter, Freitag, Fr\"ohlich,
  A{\ss}mann, Bayer, Semina, and Glazov}}]{78}
\bibinfo{author}{\bibfnamefont{J.}~\bibnamefont{Heck\"otter}},
  \bibinfo{author}{\bibfnamefont{M.}~\bibnamefont{Freitag}},
  \bibinfo{author}{\bibfnamefont{D.}~\bibnamefont{Fr\"ohlich}},
  \bibinfo{author}{\bibfnamefont{M.}~\bibnamefont{A{\ss}mann}},
  \bibinfo{author}{\bibfnamefont{M.}~\bibnamefont{Bayer}},
  \bibinfo{author}{\bibfnamefont{M.~A.} \bibnamefont{Semina}},
  \bibnamefont{and} \bibinfo{author}{\bibfnamefont{M.~M.}
  \bibnamefont{Glazov}}, \bibinfo{journal}{Phys. Rev. B}
  \textbf{\bibinfo{volume}{95}}, \bibinfo{pages}{035210}
  (\bibinfo{year}{2017}{\natexlab{a}}).

\bibitem[{\citenamefont{Zieli\'{n}ska-Raczy\'{n}ska
  et~al.}(2017)\citenamefont{Zieli\'{n}ska-Raczy\'{n}ska, Ziemkiewicz, and
  Czajkowski}}]{79}
\bibinfo{author}{\bibfnamefont{S.}~\bibnamefont{Zieli\'{n}ska-Raczy\'{n}ska}},
  \bibinfo{author}{\bibfnamefont{D.}~\bibnamefont{Ziemkiewicz}},
  \bibnamefont{and}
  \bibinfo{author}{\bibfnamefont{G.}~\bibnamefont{Czajkowski}},
  \bibinfo{journal}{Phys. Rev. B} \textbf{\bibinfo{volume}{95}},
  \bibinfo{pages}{075204} (\bibinfo{year}{2017}).

\bibitem[{\citenamefont{Schweiner
  et~al.}(2016{\natexlab{c}})\citenamefont{Schweiner, Main, Wunner, and
  Uihlein}}]{150}
\bibinfo{author}{\bibfnamefont{F.}~\bibnamefont{Schweiner}},
  \bibinfo{author}{\bibfnamefont{J.}~\bibnamefont{Main}},
  \bibinfo{author}{\bibfnamefont{G.}~\bibnamefont{Wunner}}, \bibnamefont{and}
  \bibinfo{author}{\bibfnamefont{{\relax Ch}.}~\bibnamefont{Uihlein}},
  \bibinfo{journal}{Phys. Rev. B} \textbf{\bibinfo{volume}{94}},
  \bibinfo{pages}{115201} (\bibinfo{year}{2016}{\natexlab{c}}).

\bibitem[{\citenamefont{Zieli\'{n}ska-Raczy\'{n}ska
  et~al.}(2016{\natexlab{a}})\citenamefont{Zieli\'{n}ska-Raczy\'{n}ska,
  Czajkowski, and Ziemkiewicz}}]{74}
\bibinfo{author}{\bibfnamefont{S.}~\bibnamefont{Zieli\'{n}ska-Raczy\'{n}ska}},
  \bibinfo{author}{\bibfnamefont{G.}~\bibnamefont{Czajkowski}},
  \bibnamefont{and}
  \bibinfo{author}{\bibfnamefont{D.}~\bibnamefont{Ziemkiewicz}},
  \bibinfo{journal}{Phys. Rev. B} \textbf{\bibinfo{volume}{93}},
  \bibinfo{pages}{075206} (\bibinfo{year}{2016}{\natexlab{a}}).

\bibitem[{\citenamefont{Zieli\'{n}ska-Raczy\'{n}ska
  et~al.}(2016{\natexlab{b}})\citenamefont{Zieli\'{n}ska-Raczy\'{n}ska,
  Ziemkiewicz, and Czajkowski}}]{77}
\bibinfo{author}{\bibfnamefont{S.}~\bibnamefont{Zieli\'{n}ska-Raczy\'{n}ska}},
  \bibinfo{author}{\bibfnamefont{D.}~\bibnamefont{Ziemkiewicz}},
  \bibnamefont{and}
  \bibinfo{author}{\bibfnamefont{G.}~\bibnamefont{Czajkowski}},
  \bibinfo{journal}{Phys. Rev. B} \textbf{\bibinfo{volume}{94}},
  \bibinfo{pages}{045205} (\bibinfo{year}{2016}{\natexlab{b}}).

\bibitem[{\citenamefont{Schweiner
  et~al.}(2017{\natexlab{d}})\citenamefont{Schweiner, Main, Wunner, and
  Uihlein}}]{200}
\bibinfo{author}{\bibfnamefont{F.}~\bibnamefont{Schweiner}},
  \bibinfo{author}{\bibfnamefont{J.}~\bibnamefont{Main}},
  \bibinfo{author}{\bibfnamefont{G.}~\bibnamefont{Wunner}}, \bibnamefont{and}
  \bibinfo{author}{\bibfnamefont{{\relax Ch}.}~\bibnamefont{Uihlein}},
  \bibinfo{journal}{Phys. Rev. B} \textbf{\bibinfo{volume}{95}},
  \bibinfo{pages}{195201} (\bibinfo{year}{2017}{\natexlab{d}}).

\bibitem[{\citenamefont{Schweiner
  et~al.}(2017{\natexlab{e}})\citenamefont{Schweiner, Rommel, Main, and
  Wunner}}]{275}
\bibinfo{author}{\bibfnamefont{F.}~\bibnamefont{Schweiner}},
  \bibinfo{author}{\bibfnamefont{P.}~\bibnamefont{Rommel}},
  \bibinfo{author}{\bibfnamefont{J.}~\bibnamefont{Main}}, \bibnamefont{and}
  \bibinfo{author}{\bibfnamefont{G.}~\bibnamefont{Wunner}},
  \bibinfo{journal}{Phys. Rev. B} \textbf{\bibinfo{volume}{96}},
  \bibinfo{pages}{035207} (\bibinfo{year}{2017}{\natexlab{e}}).

\bibitem[{\citenamefont{Schweiner
  et~al.}(2017{\natexlab{f}})\citenamefont{Schweiner, Main, Wunner, and
  Uihlein}}]{250}
\bibinfo{author}{\bibfnamefont{F.}~\bibnamefont{Schweiner}},
  \bibinfo{author}{\bibfnamefont{J.}~\bibnamefont{Main}},
  \bibinfo{author}{\bibfnamefont{G.}~\bibnamefont{Wunner}}, \bibnamefont{and}
  \bibinfo{author}{\bibfnamefont{{\relax Ch}.}~\bibnamefont{Uihlein}},
  \bibinfo{journal}{Phys. Rev. B}  (\bibinfo{year}{2017}{\natexlab{f}}),
  \bibinfo{note}{submitted}.

\bibitem[{\citenamefont{Schweiner
  et~al.}(2017{\natexlab{g}})\citenamefont{Schweiner, Ertl, Main, Wunner, and
  Uihlein}}]{300}
\bibinfo{author}{\bibfnamefont{F.}~\bibnamefont{Schweiner}},
  \bibinfo{author}{\bibfnamefont{J.}~\bibnamefont{Ertl}},
  \bibinfo{author}{\bibfnamefont{J.}~\bibnamefont{Main}},
  \bibinfo{author}{\bibfnamefont{G.}~\bibnamefont{Wunner}}, \bibnamefont{and}
  \bibinfo{author}{\bibfnamefont{{\relax Ch}.}~\bibnamefont{Uihlein}},
  \bibinfo{journal}{Phys. Rev. B}  (\bibinfo{year}{2017}{\natexlab{g}}),
  \bibinfo{note}{submitted}.

\bibitem[{\citenamefont{Sch\"one et~al.}(2017)\citenamefont{Sch\"one, Stolz,
  and Naka}}]{70}
\bibinfo{author}{\bibfnamefont{F.}~\bibnamefont{Sch\"one}},
  \bibinfo{author}{\bibfnamefont{H.}~\bibnamefont{Stolz}}, \bibnamefont{and}
  \bibinfo{author}{\bibfnamefont{N.}~\bibnamefont{Naka}},
  \bibinfo{journal}{Phys. Rev. B} \textbf{\bibinfo{volume}{96}},
  \bibinfo{pages}{115207} (\bibinfo{year}{2017}).

\bibitem[{\citenamefont{Kurz et~al.}(2017)\citenamefont{Kurz, Gr\"unwald, and
  Scheel}}]{94}
\bibinfo{author}{\bibfnamefont{M.}~\bibnamefont{Kurz}},
  \bibinfo{author}{\bibfnamefont{P.}~\bibnamefont{Gr\"unwald}},
  \bibnamefont{and} \bibinfo{author}{\bibfnamefont{S.}~\bibnamefont{Scheel}},
  \bibinfo{journal}{Phys. Rev. B} \textbf{\bibinfo{volume}{95}},
  \bibinfo{pages}{245205} (\bibinfo{year}{2017}).

\bibitem[{\citenamefont{Semkat et~al.}(2017)\citenamefont{Semkat, Sobkowiak,
  Sch\"one, Stolz, Koch, and Fehske}}]{95}
\bibinfo{author}{\bibfnamefont{D.}~\bibnamefont{Semkat}},
  \bibinfo{author}{\bibfnamefont{S.}~\bibnamefont{Sobkowiak}},
  \bibinfo{author}{\bibfnamefont{F.}~\bibnamefont{Sch\"one}},
  \bibinfo{author}{\bibfnamefont{H.}~\bibnamefont{Stolz}},
  \bibinfo{author}{\bibfnamefont{{\relax Th}.}~\bibnamefont{Koch}},
  \bibnamefont{and} \bibinfo{author}{\bibfnamefont{H.}~\bibnamefont{Fehske}},
  \bibinfo{journal}{arXiv:1705.08769}  (\bibinfo{year}{2017}).

\bibitem[{\citenamefont{Stielow et~al.}(2017)\citenamefont{Stielow, Scheel, and
  Kurz}}]{96}
\bibinfo{author}{\bibfnamefont{T.}~\bibnamefont{Stielow}},
  \bibinfo{author}{\bibfnamefont{S.}~\bibnamefont{Scheel}}, \bibnamefont{and}
  \bibinfo{author}{\bibfnamefont{M.}~\bibnamefont{Kurz}},
  \bibinfo{journal}{arXiv:1705.10527}  (\bibinfo{year}{2017}).

\bibitem[{\citenamefont{Heck\"otter
  et~al.}(2017{\natexlab{b}})\citenamefont{Heck\"otter, Freitag, Fr\"ohlich,
  A{\ss}mann, Bayer, Semina, and Glazov}}]{97}
\bibinfo{author}{\bibfnamefont{J.}~\bibnamefont{Heck\"otter}},
  \bibinfo{author}{\bibfnamefont{M.}~\bibnamefont{Freitag}},
  \bibinfo{author}{\bibfnamefont{D.}~\bibnamefont{Fr\"ohlich}},
  \bibinfo{author}{\bibfnamefont{M.}~\bibnamefont{A{\ss}mann}},
  \bibinfo{author}{\bibfnamefont{M.}~\bibnamefont{Bayer}},
  \bibinfo{author}{\bibfnamefont{M.~A.} \bibnamefont{Semina}},
  \bibnamefont{and} \bibinfo{author}{\bibfnamefont{M.~M.}
  \bibnamefont{Glazov}}, \bibinfo{journal}{Phys. Rev. B}
  \textbf{\bibinfo{volume}{96}}, \bibinfo{pages}{125142}
  (\bibinfo{year}{2017}{\natexlab{b}}).

\bibitem[{\citenamefont{Luttinger}(1956)}]{25}
\bibinfo{author}{\bibfnamefont{J.~M.} \bibnamefont{Luttinger}},
  \bibinfo{journal}{Phys. Rev.} \textbf{\bibinfo{volume}{102}},
  \bibinfo{pages}{1030} (\bibinfo{year}{1956}).

\bibitem[{\citenamefont{Uihlein et~al.}(1981)\citenamefont{Uihlein,
  Fr\"{o}hlich, and Kenklies}}]{7}
\bibinfo{author}{\bibfnamefont{{\relax Ch}.}~\bibnamefont{Uihlein}},
  \bibinfo{author}{\bibfnamefont{D.}~\bibnamefont{Fr\"{o}hlich}},
  \bibnamefont{and} \bibinfo{author}{\bibfnamefont{R.}~\bibnamefont{Kenklies}},
  \bibinfo{journal}{Phys. Rev. B} \textbf{\bibinfo{volume}{23}},
  \bibinfo{pages}{2731} (\bibinfo{year}{1981}).

\bibitem[{\citenamefont{Wintgen}(1987)}]{GUE5_23}
\bibinfo{author}{\bibfnamefont{D.}~\bibnamefont{Wintgen}},
  \bibinfo{journal}{Phys. Rev. Lett.} \textbf{\bibinfo{volume}{58}},
  \bibinfo{pages}{1589} (\bibinfo{year}{1987}).

\bibitem[{\citenamefont{Wintgen and Friedrich}(1987)}]{GUE1}
\bibinfo{author}{\bibfnamefont{D.}~\bibnamefont{Wintgen}} \bibnamefont{and}
  \bibinfo{author}{\bibfnamefont{H.}~\bibnamefont{Friedrich}},
  \bibinfo{journal}{Phys. Rev. A} \textbf{\bibinfo{volume}{35}},
  \bibinfo{pages}{1464(R)} (\bibinfo{year}{1987}).

\bibitem[{\citenamefont{Bohigas et~al.}(1984)\citenamefont{Bohigas, Giannoni,
  and Schmit}}]{QC_1}
\bibinfo{author}{\bibfnamefont{O.}~\bibnamefont{Bohigas}},
  \bibinfo{author}{\bibfnamefont{M.~J.} \bibnamefont{Giannoni}},
  \bibnamefont{and} \bibinfo{author}{\bibfnamefont{C.}~\bibnamefont{Schmit}},
  \bibinfo{journal}{Phys. Rev. Lett.} \textbf{\bibinfo{volume}{52}},
  \bibinfo{pages}{1} (\bibinfo{year}{1984}).

\bibitem[{\citenamefont{Brody et~al.}(1981)\citenamefont{Brody, Flores, French,
  Mello, Pandey, and Wong}}]{QC_16}
\bibinfo{author}{\bibfnamefont{T.~A.} \bibnamefont{Brody}},
  \bibinfo{author}{\bibfnamefont{J.}~\bibnamefont{Flores}},
  \bibinfo{author}{\bibfnamefont{J.~B.} \bibnamefont{French}},
  \bibinfo{author}{\bibfnamefont{P.~A.} \bibnamefont{Mello}},
  \bibinfo{author}{\bibfnamefont{A.}~\bibnamefont{Pandey}}, \bibnamefont{and}
  \bibinfo{author}{\bibfnamefont{S.~S.~M.} \bibnamefont{Wong}},
  \bibinfo{journal}{Rev. Mod. Phys.} \textbf{\bibinfo{volume}{53}},
  \bibinfo{pages}{385} (\bibinfo{year}{1981}).

\bibitem[{\citenamefont{Grosa et~al.}(2014)\citenamefont{Grosa, Legranda,
  Mortessagnea, Richalotb, and Selemanib}}]{GUE2}
\bibinfo{author}{\bibfnamefont{J.-B.} \bibnamefont{Grosa}},
  \bibinfo{author}{\bibfnamefont{O.}~\bibnamefont{Legranda}},
  \bibinfo{author}{\bibfnamefont{F.}~\bibnamefont{Mortessagnea}},
  \bibinfo{author}{\bibfnamefont{E.}~\bibnamefont{Richalotb}},
  \bibnamefont{and}
  \bibinfo{author}{\bibfnamefont{K.}~\bibnamefont{Selemanib}},
  \bibinfo{journal}{Wave Motion} \textbf{\bibinfo{volume}{51}},
  \bibinfo{pages}{664} (\bibinfo{year}{2014}).

\end{thebibliography}

\end{document}